\documentclass[aps,pra,reprint, amsmath, amssymb,superscriptaddress]{revtex4-1}
 
\usepackage{bm}
\usepackage[retainorgcmds]{IEEEtrantools}
\usepackage{graphicx}
\usepackage{mathrsfs}
\usepackage{amsmath}
\usepackage{amsfonts}
\usepackage{amssymb}
\usepackage{color}
\usepackage{times,txfonts}
\usepackage{nicefrac}
\usepackage{ragged2e}
\usepackage{tikz}

\usepackage[colorlinks=true,linkcolor=blue,urlcolor=blue,citecolor=blue,pdfusetitle]{hyperref}

\usepackage{blindtext}

\usepackage[framemethod=TikZ]{mdframed}
\usepackage[T1]{fontenc}
\mdfdefinestyle{MyFrame}{%
    linecolor=black,
    outerlinewidth=1pt,
    roundcorner=0pt,
    innertopmargin=0.77\baselineskip,
    innerbottommargin=0.75\baselineskip,
    innerrightmargin=14pt,
    innerleftmargin=14pt,
    backgroundcolor=gray!20!white}
\mdfdefinestyle{MyFrameTitled}{%
    linecolor=black,
    outerlinewidth=1pt,
    roundcorner=0pt,
    innertopmargin=0.77\baselineskip,
    innerbottommargin=0.75\baselineskip,
    innerrightmargin=14pt,
    innerleftmargin=14pt,
    backgroundcolor=gray!20!white,
     frametitlebackgroundcolor   =black!15,
    frametitlerule =true,
}

\newcommand{\ket}[1]{\vert #1 \rangle}
\newcommand{\bra}[1]{\langle #1 \vert}
\newcommand{\braket}[2]{\langle #1 \vert #2 \rangle}

\newcommand{\tr}[1]{\mathrm{tr}\left[#1\right]}

\usepackage{amssymb}% http://ctan.org/pkg/amssymb
\usepackage{pifont}% http://ctan.org/pkg/pifont

\begin{document}

\title{Diverging current fluctuations in critical Kerr resonators}

\date{\today}

\author{Michael J. Kewming}
\email{kewmingm@tcd.ie}
\affiliation{School of Physics, Trinity College Dublin, College Green, Dublin 2, Ireland}

\author{Mark T. Mitchison}
\email{mark.mitchison@tcd.ie}
\affiliation{School of Physics, Trinity College Dublin, College Green, Dublin 2, Ireland}

\author{Gabriel T. Landi}
\email{gtlandi@gmail.com}
\affiliation{School of Physics, Trinity College Dublin, College Green, Dublin 2, Ireland}
\affiliation{Instituto de F\'isica da Universidade de S\~ao Paulo,  05314-970 S\~ao Paulo, Brazil.}

\begin{abstract}
The parametrically pumped Kerr model describes a driven-dissipative nonlinear cavity, whose nonequilibrium phase diagram features both continuous and discontinuous quantum phase transitions. We consider the consequences of these critical phenomena for the fluctuations of the photocurrent obtained via continuous weak measurements on the cavity. Considering both direct photodetection and homodyne detection schemes, we find that the current fluctuations diverge exponentially at the discontinuous phase transition. However, we find strikingly different current fluctuations for these two detection schemes near the continuous transition, a behaviour which is explained by the complementary information revealed by measurements in different bases. To obtain these results, we develop formulas to efficiently compute the diffusion coefficient---which characterises the long-time current fluctuations---directly from the quantum master equation, thus connecting the formalisms of full counting statistics and stochastic quantum trajectories. Our findings highlight the rich features of current fluctuations near nonequilibrium phase transitions in quantum-optical systems.
\end{abstract}

\maketitle{}

% \begin{mdframed}[style=MyFrameTitled, frametitle = To-Do list]
% Goal: understand fluctuations in the output current, and how they relate to continuous and discontinuous transitions. 
% \begin{itemize}
%     \item {\bf Stochastic simulations:} have been studied in Ref.~\cite{Bartolo2017}.
%     \begin{itemize}
%         \item Homodyne trajectories on phase space, with the Wigner function on top. 
%         \item Photo-detection trajectories and the strange drop at $t = 3$. 
%         \item How to guarantee that the simulations are converging properly? (model is tricky!)
%     \end{itemize}
%     \item {\bf Understand the fluctuations:}
%     \begin{itemize}
%         \item How to plot/visualise excited states of a Liouvillian?
%         \item Phase-space perspective using the Wigner-Weyl representation.
%     \end{itemize}
%     \item {\bf Analyze $\mathcal{D}$ for the homodyne current} \cmark
%     \begin{itemize}
%         \item Check if plateau has any relation to $\lambda_1$. No it is not. \cmark
%     \end{itemize}
%     \item Analyze $\lambda_1$ in the $(G,\Delta)$ plane. 
%     \item Study $S(\omega)$ (fluctuations at non-zero frequency).
%     \item Understand in which transitions there is spontaneous symmetry breaking. Check~\cite{Bartolo2016,Minganti2018,Minganti2019}
% \end{itemize}

% \end{mdframed}

\section{\label{sec:int}Introduction}

One of the core tenets of quantum theory is the fundamentally random character of measurement. This randomness has deep consequences for the foundations of quantum mechanics~\cite{Schlosshauer_decoherence_2007}, but  is also practically important for any experimenter trying to unravel the dynamics of a quantum system. A pertinent example consists of systems subject to continuous weak measurement, which are now routinely studied in quantum optics~\cite{Nagourney1986,Sauter1986,Bergquist1986} and mesoscopic physics~\cite{Vijay2011, Minev2019} experiments. The resulting measurement outcomes are classical stochastic processes that we refer to here as \textit{measurement currents}: they are the random stream of clicks in photodetectors, the diffusive wandering of homodyne signals, and the fluctuating currents through quantum point contacts. % references here?
It is from these currents that properties of continuously measured quantum systems are inferred. However, the statistical character of the observed current may differ significantly depending on the underlying quantum state. This difference can be rich and illuminating, especially in the vicinity of phase transitions~\cite{Nguyen_2020,fiore_current_2021}.

A nonequilibrium phase transition~\cite{Marro_1999} occurs when a system undergoes a sudden change in its nonequilibrium steady state (NESS). Such transitions can be described as either continuous or discontinuous, corresponding to the behavior of the order parameter across the critical point. Any NESS is characterized by currents of quantities such as particles and energy, and these currents may become singular near the critical point. A current of particular interest is the entropy production rate, which is always nonzero in a NESS by definition. There have been several studies of classical \cite{Tom__2012, Shim_2016, Crochik_2005, Zhang_2016, Noa_2019, Herpich_2018, Herpich_2019} and quantum-optical models \cite{Brunelli2018, Goes_2020}, showing that the average entropy production rate at criticality can distinguish the nature of a nonequilibrium phase transition. Specifically, at continuous transitions it is divergent but continuous~\cite{Tom__2012, Shim_2016, Crochik_2005}, while at discontinuous transitions it becomes discontinuous \cite{Zhang_2016, Herpich_2018}. 
Similar results are also expected for other types of output currents.
More recently, it has been realised that current  fluctuations  exhibit dramatically different scaling near the critical point of continuous and discontinuous transitions in the classical regime~\cite{Jordan2004,Schaller2010,Nguyen_2020,fiore_current_2021}. 
However, to date there is little commentary on the fluctuations of observed currents in a purely quantum-mechanical phase transition. 

Here, we address this problem in the context of the parametrically pumped Kerr (PPK) model, which describes a lossy, nonlinear optical cavity subject to a two-photon drive, and is a paradigmatic example of optical bistability \cite{drummond_quantum_1980, Wolinsky_1988}. The competition between dissipation and the parametric drive leads to the system becoming critical~\cite{hines_quantum_2005,meaney_quantum_2014,bartolo_exact_2016, Minganti_2018, kewming_quantum_2020, Zhang_2021}, with a nonequilibrium phase diagram featuring both continuous and discontinuous transitions.  Criticality arises here in the limit of weak nonlinearity, $U\to 0$, so the parameter $1/U$ plays the role of volume in the conventional thermodynamic limit, analogous to other finite-component phase transitions~\cite{Ashhab2013, Hwang2015,Carmichael2015,Felicetti2020}. 

We assume that the photons lost from the cavity are continuously monitored, generating a discrete photocurrent or diffusive homodyne current when mixed with a local oscillator. Naturally, the onset of criticality should present itself in the fluctuations of the measured current, a fact which has been explored in other nonequilibrium phase transition models~\cite{Nguyen_2020,fiore_current_2021}. 
However, a key novelty of our setup is the freedom to measure the system in different bases (e.g.~via direct or homodyne detection of emitted photons), giving rise to very different conditional dynamics driven by quantum measurement backaction. To quantify the current fluctuations near criticality, we consider the steady-state diffusion coefficient $\mathcal{D}$, which characterises the asymptotic dispersion of the integrated measurement current. We employ tools from the theory of full counting statistics (FCS)~\cite{levitov_charge_1993,esposito_nonequilibrium_2009,Schaller2014} to develop formulas for $\mathcal{D}$, which are numerically efficient to evaluate for both measurement schemes. Doing so also serves to tighten the connection between FCS, prevalent in condensed-matter physics, and the quantum continuous measurement approach used in quantum optics~\cite{wiseman_quantum_2009, Jacobs2014}. 

Using these tools, we examine the finite-size scaling of the current fluctuations close to criticality in the PPK model. We find that fluctuations diverge exponentially with $1/U$ at the discontinuous phase transition for both measurement schemes, in accordance with results from classical systems~\cite{Nguyen_2020,fiore_current_2021}. Interestingly, however, the fluctuations of the homodyne current also diverge exponentially at the continuous phase transition, in contrast to the photocurrent which diverges algebraically in accordance with the expectation from classical models~\cite{Nguyen2018}. As we explain, this discrepancy is due to the nature of the measurement and the information that each reveals about the NESS: Exponential divergence will occur when the underlying bistable property of the system produces a telegraphic-like switching in the observed current.

Before progressing onto the main body of this manuscript, we briefly outline its structure. Initially in Sec.~\ref{sec:The model}, we summarise the characteristics of the PPK model, detailing its continuous and discontinuous phase transitions. The former can be predicted semi-classically, while the latter cannot, thus being purely quantum mechanical in nature. In Sec.~\ref{sec:Quantum Trajectories} we introduce the quantum trajectories formalism for both photodetection and homodyne detection and use it to study the dynamics of the PPK model close to criticality. In Sec.~\ref{sec: measurements}, we detail our approach to compute the power spectrum $S(\omega)$ and subsequently the diffusion coefficient $\mathcal{D}$, which measures the divergent current fluctuations close to criticality. 
% However, we show that that in a certain region of parameter space, this divergence is algebraic in the system parameters---corresponding to the continuous phase transition----while, in another it is exponential---corresponding to the discontinuous phase transition. 
Finally, in Sec.~\ref{sec:D_results} we study the nature of this divergence and show that it crucially depends on the employed measurement scheme. To explain both the exponential divergences and this differing behaviour we show that the rate of divergence in the PPK at the discontinuous phase transitions is determined by the average tunnelling time between the metastable fixed states. Our conclusions are presented in Sec.~\ref{sec:conclusions}.

%%%%%%%%%%%%%%%%%%%%%%%%
\section{The parametrically driven Kerr model}
\label{sec:The model}

\begin{figure*}
    \centering
    \includegraphics[width=\textwidth]{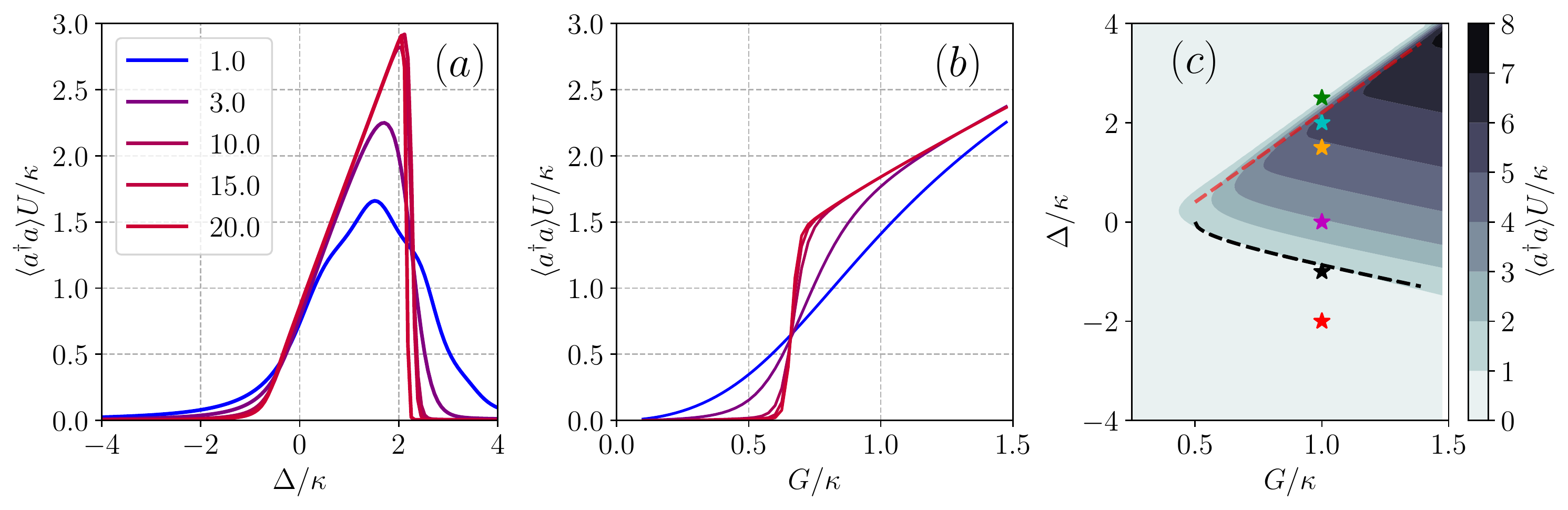}
    \caption{Phase diagram of the PPK model described by Eqs.~\eqref{H} and~\eqref{M}.
    (a) Steady-state average photon number $\langle a^\dagger a \rangle U/\kappa$ as a function of $\Delta/\kappa$, for different values of $\kappa/U$ (legend) and fixed $G/\kappa = 1.0$. For increasing values of $\kappa/U$ the discontinuous transition becomes sharper around $\Delta_{d} \sim 2$.
    The model becomes critical in the limit $U/\kappa\to 0$, undergoing a continuous transition at $\Delta <0$ and a discontinuous one at $\Delta >0$. 
    (b) A plot of $\langle a^\dagger a \rangle U/\kappa$ as a function of $G/\kappa$ for the same values of $\kappa/U$ (legend in a)), with $\Delta/\kappa = 1$, showcasing the discontinuous nature of the transition when $\Delta >0$. For increasing $\kappa/U$ the transition again becomes sharper and approaches a discontinuous step.
    (c)  $\langle a^\dagger a \rangle U/\kappa$ in the $(G/\kappa,\Delta/\kappa)$ plane for  $\kappa/U = 10$, sometimes called the instability tongue. The lower black-dashed line denotes the continuous transition, determined analytically from the semi-classical theory as $\Delta_c = -\sqrt{G^2 - \kappa^{2}/4}$. The upper red-dashed line indicates the discontinuous transition. The asterisks correspond to different choices of $(G/\kappa,\Delta/\kappa)$ which are used below in Fig.~\ref{fig:wigner1}.
    }
    \label{fig:phase_diagram}
\end{figure*}

In this section we review the known critical properties of the PPK model. 
The model consists of a single bosonic cavity mode of frequency $\omega_{a}$ containing a Kerr non-linearity with strength $U$, proportional to the third-order nonlinear susceptibility.
The cavity is parametrically driven by a pump mode at the frequency $\omega_{p}\sim 2\omega_a$, which creates two excitations in the cavity mode per absorbed photon. Assuming a sufficiently large separation of timescales between the cavity and pump mode dynamics, we can adiabatically eliminate the pump and describe the dynamics of the cavity mode via the interaction-picture Hamiltonian $(\hbar=1)$~\cite{drummond_quantum_1980, Wolinsky_1988,roberts_driven-dissipative_2019}
\begin{equation}
\label{H}
    \hat{H} = - \Delta \hat{a}^\dagger \hat{a} + \frac{U}{2} \hat{a}^{\dagger 2} \hat{a}^{2} + \frac{G}{2} (\hat{a}^{\dagger 2} + \hat{a}^{2})\,,
\end{equation}
where $\Delta = \omega_{a} - \omega_{p}/2$ is the detuning of the cavity from the pump, $G$ corresponds to the strength of the two-photon parametric pump, and $\hat{a}$ $(\hat{a}^{\dagger})$ is the annihilation (creation) operator of the mode satisfying the commutation relation $[\hat{a}, \hat{a}^{\dagger}] = 1$. 
We further assume the cavity is subject to single-photon losses, with a loss rate $\kappa$. As such, it can be described by the Born-Markov quantum master equation
\begin{equation}
\label{M}
    \frac{d\hat{\rho}}{dt} = \mathcal{L}(\hat{\rho}) = -i [\hat{H},\hat{\rho}] + \kappa \Big[\hat{a} \hat{\rho} \hat{a}^\dagger - \frac{1}{2} \{\hat{a}^\dagger \hat{a}, \hat{\rho}\}\Big].
\end{equation} 
Throughout, we will generally use $\kappa$ as our basic scale, plotting all other parameters in units of $\kappa=1$.

The general solution to this time-independent master equation is 
$\hat{\rho}(t) = e^{\mathcal{L}t}\hat{\rho}(0)$.
In the long-time limit, the system will generally tend to a unique nonequilibrium steady state, $\hat{\rho}_{ss}$, which is the solution of $\mathcal{L}(\hat{\rho}_{ss}) = 0$. This can be computed analytically using either the generalized P function method~\cite{drummond_generalised_1980, meaney_quantum_2014, bartolo_exact_2016} or the coherent quantum absorber approach~\cite{stannigel_driven-dissipative_2012,roberts_driven-dissipative_2019}. The NESS may undergo both a continuous or a discontinuous phase transition as the parameters of the Hamiltonian are varied. Technically speaking, the model only becomes critical (in the sense that quantities become non-analytic) in the limit $U \to 0$~\cite{bartolo_exact_2016}. 
We shall therefore analyze the results for different values of $1/U$ in order to characterize their scaling as $U\to 0$. 
The association of $U\to 0$ with a thermodynamic limit was discussed in detail in Ref.~\cite{Vicentini_2018}. There the authors showed that, starting from a many-body lattice model with similar interactions plus a hopping term, performing a mean-field approximation leads to the same Hamiltonian as in Eq.~\eqref{H} but with $U$ scaled as $U/N$, where $N$ is the actual number of particles in the lattice.
% In fact, as shown in~\cite{Vicentini_2018}, 
% Eq.~\eqref{H} can be viewed as 
% the parameter $1/U$ would play the role of the particle number in a many-body lattice model with similar interactions and additional tight-biding hoping. Thus the limit $1/U \to \infty$ is analogous to the thermodynamic limit. 

Typically the nature of the phase transitions arising in the PPK model is analyzed using the semi-classical equations of motion, where Heisenberg-picture operators $\hat{a}\, (\hat{a}^{\dagger})$ are replaced by complex variables $\alpha\, (\alpha^{*})$.
% Th at $\Delta < 0$, can actually be well predicted by a semi-classical theory,
From the semi-classical analysis, one can show~\cite{meaney_quantum_2014} that above a critical driving, the system undergoes a pitchfork bifurcation with an unstable fixed point at $\alpha = 0$ and stable fixed points at $\alpha = \pm \alpha_0$, with $\alpha_0 = \sqrt{n_0}e^{i\phi_0}$ and
\begin{align}
\label{eq:semi-classical}
    n_{0} = \frac{\Delta}{U} +\sqrt{\frac{G^{2} - \kappa^{2}/2}{U^{2}}}\,, \quad \phi_{0} = \frac{1}{2}\arcsin \left(-\frac{\kappa}{2 G}\right)\,.
\end{align}
Imposing that $n_0>0$ determines the critical detuning $\Delta_c = -\sqrt{G^2 - \kappa^{2}/4}$ of this pitchfork bifurcation.
For large driving $G$, the steady state is very well approximated by an equal mixture of positive and negative coherent states
\begin{align}
\label{eq:ss}
    \hat{\rho} \approx \frac{1}{2}\left(\ket{\alpha_0}\bra{\alpha_0} + \ket{-\alpha_0}\bra{-\alpha_0}\right)\,.
\end{align}

% While this phase transition has been the subject of numerous research studies as described in the Introduction of this manuscript, there exists an additional phase transition which has received considerably less attention. As $\Delta$ increases above the critical detuning $\Delta > \Delta_{c}$, the bimodal stable fixed point become trimodal until above some critical detuning $\Delta_d$ where the bimodal fixed points become unstable resulting in a sudden phase transition back to the squeezed vacuum. 
% This transition does not correspond to any semi-classical bifurcation, and thus is purely quantum mechanical in nature \cite{meaney_quantum_2014}. 

The PPK model also features a discontinuous transition occurring at a positive detuning $\Delta_d=2$, which has been studied previously \cite{meaney_quantum_2014, Bartolo_2016, Minganti_2018}.
To see this, 
% To better understand the phase transitions,
we plot in Fig.~\ref{fig:phase_diagram} the numerically computed steady-state mean photon number $\langle \hat{a}^{\dagger}\hat{a}\rangle U/\kappa$ --- scaled by $\kappa/U$ --- for different choices of $\Delta/\kappa$ and $G/\kappa$.
As clearly seen in Fig.~\ref{fig:phase_diagram}(a), $\langle \hat{a}^\dagger \hat{a}\rangle$ undergoes a continuous transition at $\Delta_c<0$ and then a discontinuous jump at $\Delta_{d}$, where both transitions become increasingly sharp for increasing values of $\kappa/U$. The discontinuous nature of the second transition is further revealed by plotting $\langle \hat{a}^{\dagger}\hat{a}\rangle$ as a function of the parametric drive strength $G/\kappa$ for increasing values of $\kappa/U$, with fixed $\Delta/\kappa = 1$ (Fig.~\ref{fig:phase_diagram}(b)): At the critical driving, the mean occupation of the system suddenly jumps from zero to a finite value. Both transitions depend sensitively on the interplay between the driving $G$ and the detuning $\Delta$. We can study this interplay by constructing the phase diagram of the mean photon number in the $(G,\Delta)$ plane, as shown in Fig.~\ref{fig:phase_diagram}(c).
The critical region corresponds to $G > \kappa/2$. For $\Delta < 0$ the transition is continuous and the corresponding critical line $\Delta_c = -\sqrt{G^2 - \kappa^{2}/4}$ is shown by the black dashed curve. Conversely, for $\Delta >0$ the transition is discontinuous and cannot be captured by the semi-classical theory described above. 
The red-dashed line in Fig.~\ref{fig:phase_diagram}(c) indicates the approximate location of the discontinuous phase transition critical line $\Delta_d(G)$, as inferred from the numerical results.
Similar behaviour exists in Duffing resonators \cite{Lifshitz, Dykman_1998} which exhibit regions of instability, sometimes referred to as instability `tongues' in direct analogy with the critical (shaded) region in Fig.~\ref{fig:phase_diagram}(c). The instability tongue and similar amplitude curves as Fig.~\ref{fig:phase_diagram}(a) and (b) have been observed experimentally in both micromechanical \cite{Turner_1998} and electromechanical \cite{Mahboob_2008, Mahboob_2014} systems.
% For each coloured Wigner function plotted in Fig.~\ref{fig:wigner1}, we have marked, with an asterisk, the corresponding $(G,U)$ parameters in Fig.\ref{fig:phase_diagram}(c) clearly showing the discontinuous phase transition.

\begin{figure}
    \centering
    \includegraphics[width=0.45\textwidth]{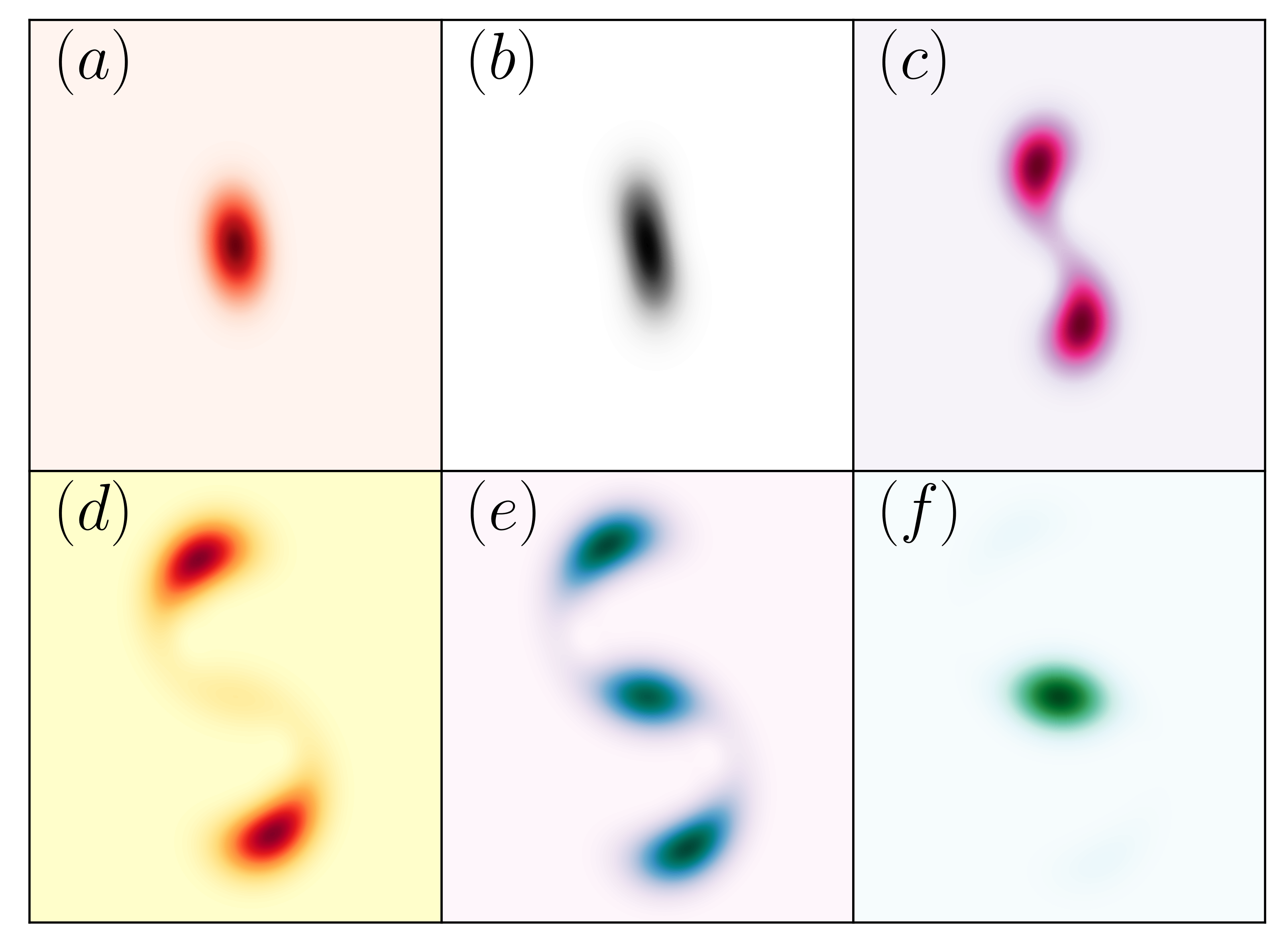}
    \caption{Steady-state Wigner function $W(x,p)$ at the different asterisks shown in Fig.~\ref{fig:phase_diagram}(c) (from bottom to top) corresponding to $\Delta/\kappa = (-2, -1, 0, 1.5, 2, 2.5)$, with fixed $G/\kappa=1$ and $U/\kappa = 1/3$. In all cases, the Wigner function is non-negative (the colorbars are omitted for visibility).
    }
    \label{fig:wigner1}
\end{figure}

We can further visualise the nature of these phase transitions by studying the steady-state Wigner function $W(x, p)$, where $(x,p)$ are the field quadratures. Fig.~\ref{fig:wigner1} shows example plots of  $W(x,p)$ for the points marked in Fig.~\ref{fig:phase_diagram}(c). For $\Delta \ll 0$, the steady state is roughly a squeezed vacuum, which then undergoes the continuous phase transition to the bimodal state for increasing detuning. We can clearly see the trimodal nature of the NESS arising for $\Delta \sim \Delta_{d}$ and then suddenly becoming a squeezed vacuum again, once the discontinuous phase boundary is crossed. A similar trimodal structure is observed in the Duffing resonator \cite{Dykman_1998, Yamaji_2022}, thus further establishing the analogy with the PPK model.
% To shed further light on the emergence of the discontinuous transition, we further plot in Fig.~\ref{fig:wigner2} the Wigner function fixed at the critical detuning $\Delta_d$, but now vary the strength of the Kerr non-linearity for different values of $1/U$.
% As can be seen, increasing $1/U$ causes the 3 the trimodal structure of the Wigner function to stretch along the vertical axis. In addition, it also causes the relative contribution of the middle spot to decrease. 
% In fact, the main effect of increasing $1/U$ is to cause said relative contribution to become strongly dependent on $\Delta$, and thus increasing the discontinuity of the transition from the bimodal structure to the squeezed vacuum.

% \begin{figure}
%     \centering
%     \includegraphics[width=0.45\textwidth]{wigner2.png}
%     \caption{Wigner function of the discontinuous transition point for varying $U$. 
%     Each plot shows the steady state Wigner function $W(x,p)$ for different values of $1/U = 1,2,3,5,7,10$, with fixed $G = \kappa = 1$ and $\Delta = \Delta_{d} \approx 2$. Image (c) is the same as Fig.~\ref{fig:wigner1}(e).
%     }
%     \label{fig:wigner2}
% \end{figure}

\section{Quantum trajectories and currents}
\label{sec:Quantum Trajectories}

So far we have discussed the steady state of the PPK model. We are now interested in studying its behaviour when subject to continuous measurements. Continuous measurement of a quantum system conditions the dynamics on the measurement outcomes. Given that the outcome of these measurements are random, the evolution of the state follows stochastic quantum trajectories, the theory of which is well developed  \cite{wiseman_quantum_2009,Jacobs2014}. Ultimately we will be concerned with the measurement current $I(t)$ and its fluctuations, characterised by the diffusion coefficient $\mathcal{D}$. Before proceeding to this description, however, it is prudent to first summarise the method of quantum trajectories for two commonly used measurement schemes:  photodetection and homodyne detection. The measurement outcomes of these two schemes reveal differing characteristics about the quantum state and its dynamics in the critical region.

\subsection{Photodetection}

For perfect photodetection---where the detector efficiency $\eta=1$ so that every emitted photon is detected---the evolution of the density operator is governed by an It\^o stochastic differential equation
\begin{equation}
    d \hat{\rho}_{\rm PD} = -\mathcal{H}\left[i \hat{H} + \frac{\kappa}{2}\hat{a}^{\dagger}\hat{a}\right]\hat{\rho}_{\rm PD}dt +  dN(t)\mathcal{G}\left[\hat{a}\right]\hat{\rho}_{\rm PD}\,,
\end{equation}
where $dN(t)$ is a stochastic Poisson increment satisfying $dN(t)^{2} = dN(t)$ and $\hat{\rho}_{\rm PD}$ indicates that the density operator has been conditioned on the photodetections. The two superoperators above are defined by
\begin{equation}
    \mathcal{G}[\hat{A}]\hat{\rho} = \frac{\hat{A}\hat{\rho} \hat{A}^{\dagger}}{\langle \hat{A}^{\dagger}\hat{A}\rangle } - \hat{\rho}\,, \quad 
    \mathcal{H}[\hat{A}]\hat{\rho} = \hat{A}\hat{\rho} + \hat{\rho}\hat{A}^{\dagger} - \langle \hat{A}\hat{\rho} + \hat{\rho}\hat{A}^{\dagger} \rangle \,.
\end{equation}
The observed photocurrent $I_{\rm PD}(t) = dN(t)/dt$ is a series of $\delta$-like peaks, corresponding to photons being registered as `clicks' in the detector. Each click induces a discrete quantum jump described by the superoperator $\mathcal{G}[\hat{a}]$. The average jump rate is given by ${\rm E}[dN(t)]=\kappa \tr{\hat{a} \hat{\rho} \hat{a}^\dagger}dt$, where $\rm E[\bullet]$ denotes a classical average over stochastic trajectories.
%Thus probability for this jump to occur is $P(dN(t) = 1) = \kappa \tr{\hat{a} \hat{\rho} \hat{a}^\dagger}$.
% Moreover, this evolution would only reveal information about the photon number of the cavity at any given time and cannot resolve any of the quadrature $\langle \hat{x} \rangle = \langle \hat{q}_{\theta=0} \rangle$ or $\langle \hat{p} \rangle = \langle \hat{q}_{\theta=\pi/2} \rangle$.

\begin{figure}
    \centering
    \includegraphics[width=\columnwidth]{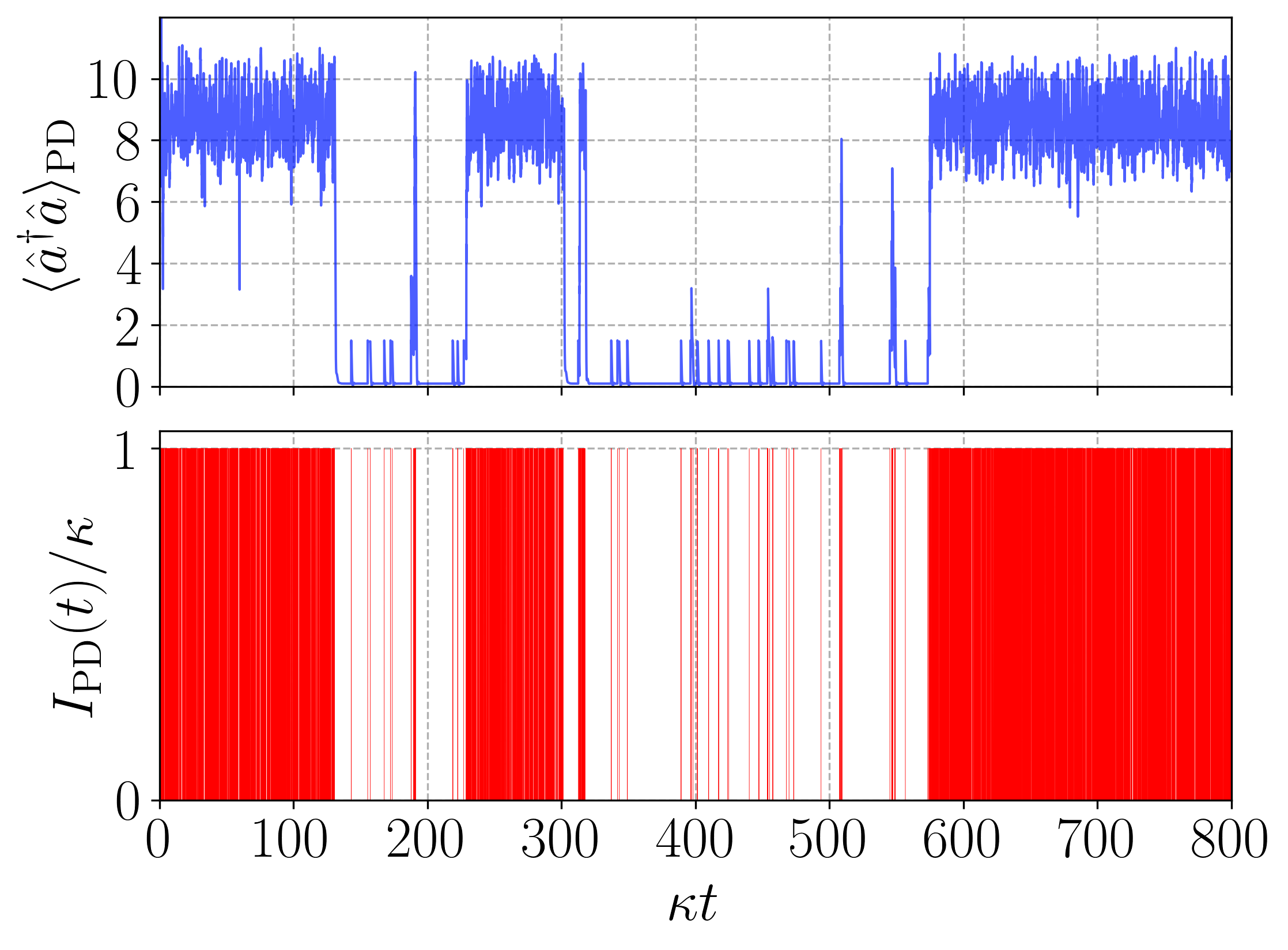}
    \caption{Quantum trajectories for direct photodetection.
    The parameters are fixed at $G/\kappa=1$, $U/\kappa=1/3$,  $\Delta/\kappa =2$, corresponding to the onset of the discontinuous transition (Fig.~\ref{fig:wigner1}(e)). 
    (Top) Conditional mean photon number $\langle \hat{a}^{\dagger}\hat{a}\rangle_{\rm PD}$  as a function of time during a single trajectory.
    (Bottom) Observed photocurrent $I_{\rm PD}(t)$, corresponding to a series of Dirac delta functions at each detection event. 
    The rate of clicks in the detector reduces dramatically whenever the system is in the central lobe of Fig.~\ref{fig:wigner1}(e). }
    \label{fig:photodetection}
\end{figure}

In the top panel of Fig.~\ref{fig:photodetection}  we plot the conditional mean photon number $\langle \hat{a}^{\dagger}\hat{a}\rangle_{\rm PD}$ of the cavity, alongside the conditional variance $\Delta ( \hat{a}^{\dagger}\hat{a})_{\rm PD} = \langle (\hat{a}^{\dagger}\hat{a})^{2}\rangle_{\rm PD} - \langle \hat{a}^{\dagger}\hat{a}\rangle_{\rm PD}^{2}$, when the system is very close to the discontinuous phase transition. We see that the cavity flips stochastically between two very different states: one where the cavity is effectively empty and another where it is highly populated. These correspond to the inner and outer fixed points of the trimodal Wigner function shown in Fig.~\ref{fig:wigner1}(e). % Notice that the variance scales as $\Delta(\hat{a}^{\dagger}\hat{a})_{\rm Hom}\sim \langle \hat{a}^{\dagger}\hat{a}\rangle_{\rm Hom}^{2}$. 
In the lower panel  of Fig.~\ref{fig:photodetection} we also plot the observed time-series of the photon current $I_{\rm PD}$, which shows that the rate of detection events rapidly changes between 
% either side of the phase transition:
an active phase, with a large number of emissions, and an inactive phase with essentially no clicks.
Thus, from the photon current it is possible to directly observe the system switching between its two configurations near the critical point.
Crucially, however, photodetection cannot distinguish between the two outer fixed points because it measures the amplitude and not the phase of the output field.

\begin{figure*}
    \centering
    \includegraphics[width=\textwidth]{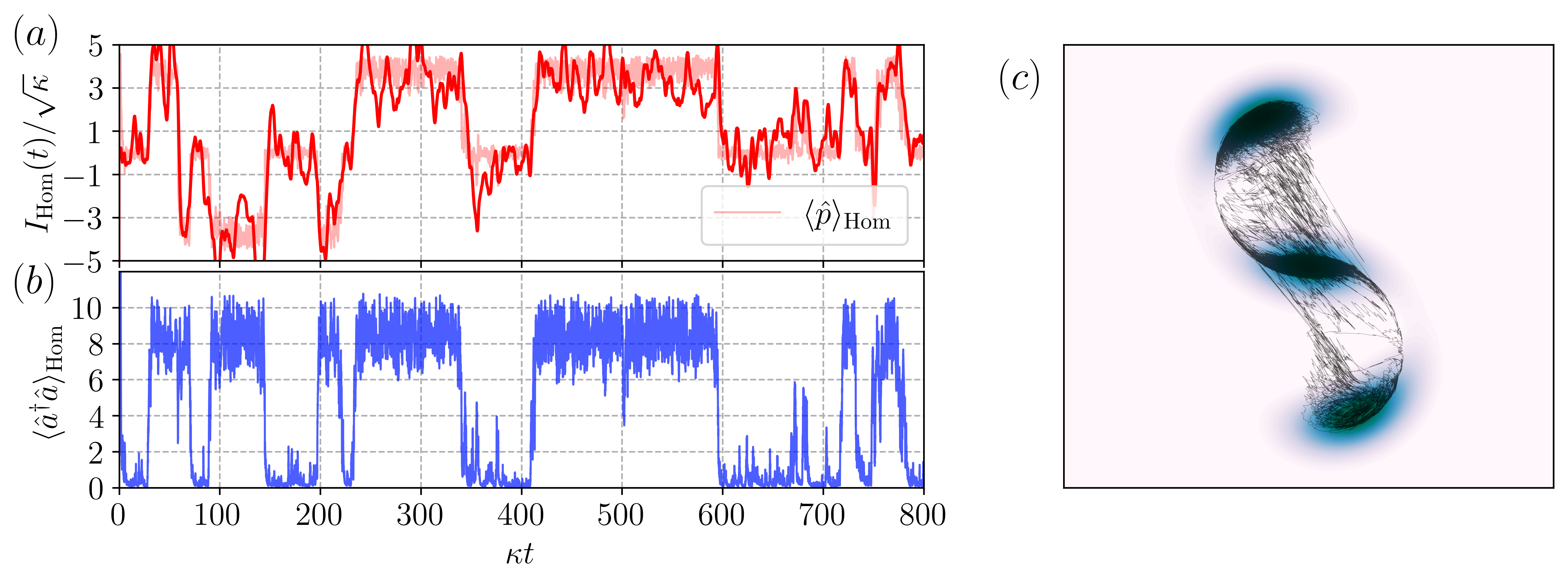}
    \caption{Quantum trajectories for homodyne detection along $\hat{p}$, for the same parameters as in Fig.~\ref{fig:photodetection}. 
    The curves showcase the  tunnelling dynamics of the system, between three metastable states.
    (a) Homodyne current $I_{\rm Hom}(t)$ (dark red) and 
    underlying unconditional moment $\langle \hat{p}\rangle_{\rm Hom}$ (light red); 
    the former was processed by a low-pass filter to remove high frequency noise.    
    % Near the phase transition we can clearly see the system tunnelling between the three metastable states. The measurement localises the state along the $\hat{p}$ axis thus collapsing the Cat-state into a displaced coherent state. 
    (b) Corresponding  mean photon number, which cannot distinguish between the outer lobes, but shows a sudden reduction when the system tunnels to the central lobe. (c) Quantum trajectories of $(\langle \hat{x} \rangle_{\rm Hom}, \langle \hat{p} \rangle_{\rm Hom})$, overlaid on the associated Wigner function of Fig.~\ref{fig:wigner1}(e). }
    \label{fig:homodyne}
\end{figure*}

\subsection{Homodyne detection}

In the homodyne (diffusive) unravelling, one measures instead a generic quadrature of the form $\hat{q}_{\theta} =  \hat{a}e^{-i \theta} + \hat{a}^{\dagger}e^{i\theta}$  
(the two directions $x$ and $p$ in the Wigner function plots of Fig.~\ref{fig:wigner1} correspond to $\hat{x} = \hat{q}_{\theta=0} = \hat{a} + \hat{a}^\dagger$ and $\hat{p} = \hat{q}_{\theta=\pi/2} = i(\hat{a}^\dagger-\hat{a})$). 
The It\^o stochastic master equation describing the evolution of the density operator subject to perfect homodyne detection along the quadrature $\hat{q}_{\theta}$ is~\cite{wiseman_quantum_2009}
\begin{align}
\label{eq:homodyne}
  d \hat{\rho}_{\rm Hom} = \left(- i [\hat{H}, \hat{\rho}_{\rm Hom}] +  \kappa D[\hat{a}e^{i\theta}]\hat{\rho}_{\rm Hom}\right) dt \nonumber \\ +\sqrt{\kappa}\mathcal{H}\left[\hat{a}e^{i\theta}\right]\hat{\rho}_{\rm Hom}dW(t)\,,
\end{align}
where $D[\hat{a}]\rho {=} \hat{a} \hat{\rho} \hat{a}^\dagger - \frac{1}{2} \{\hat{a}^\dagger \hat{a},\hat{\rho}\}$ and
$dW(t)$ is a Wiener increment 
with  $\mathrm{E}[dW(t)] {=} 0$, and a variance $\mathrm{Var}[dW(t)] {=} dt$. 
The current $I_{\rm Hom}(t)$---after the background oscillator has been subtracted---is given by 
\begin{equation}
    I_{\rm Hom}(t) = \sqrt{\kappa}\,\tr{\hat{q}_{\theta}\hat{\rho}_{\rm Hom}} + \xi(t)\,,
\end{equation}
where $\xi(t) = dW(t)/dt$. 
In the examples considered below, we will focus on $\theta = \pi/2$
% be concerned with measuring the $\hat{p} = i\hat{a} - i \hat{a}^{\dagger}$ quadrature, i.e. $\theta = \pi/2$, 
as it corresponds to the separation axis of the bifurcation depicted in Fig.~\ref{fig:wigner1}.

As an example, Fig.~\ref{fig:homodyne} shows the conditional evolution under homodyne detection along $\hat{p}$, at the same configuration as Fig.~\ref{fig:wigner1}(e) (close to the discontinuous phase transition). 
In Fig.~\ref{fig:homodyne}(a), the homodyne current $I_{\rm Hom}$ (after passing through a low-pass filter) is plotted together with the conditional moment $\langle \hat{p}_{\rm Hom}\rangle$. 
As can be seen, both clearly follow a tristable behaviour, with the system tunnelling between each of the steady-state fixed points.
This can also be visualised in Fig.~\ref{fig:homodyne}(c), where we plot the stochastic trajectories of the system on top of the steady-state Wigner function.
For comparison, the average photon number $\langle a^\dagger a \rangle_{\rm Hom}$ is also plotted in Fig.~\ref{fig:homodyne}(b).
When $I_{\rm Hom} \neq 0$, irrespective of whether it is positive or negative, the system will be in an active state where $\langle a^\dagger a \rangle_{\rm Hom}$ is non-zero.
This underscores a key difference between photodetection and homodyne detection: the latter can resolve all three steady-state fixed points, while photodetection can only distinguish the outer pair from the inner one. As we will show, this leads to very different fluctuations of the two measurement currents across the critical region. 

\section{Quantifying current fluctuations via the diffusion coefficient}
\label{sec: measurements}

% One of the primary signatures of a discontinuous phase transition is the exponential divergence of measured current fluctuations $I(t)$. 
The fluctuations of the output currents can be quantified via the diffusion coefficient $\mathcal{D}$, which is a central quantity in the theory of full counting statistics (FCS)~\cite{levitov_charge_1993,esposito_nonequilibrium_2009,fiore_current_2021}. 
% In this section we will show that at in the PPK at the discontinuous phase transition, the fluctuations diverge exponentially with increasing $1/U$.
%In this section, we will show that the fluctuations in the measurement currents---the photocurrent $I_{\rm PD}(t)$ and the homodyne current $I_{\rm Hom}(t)$---diverge exponentially with increasing $1/U$ at the discontinuous phase transition. Moreover, we will investigate the properties of these fluctuations and show that they depend intrinsically on the nature of the measured currents.
In this section, we provide the precise definition of $\mathcal{D}$ and develop efficient expressions for computing it. These will be used in the following section to investigate the scaling of current fluctuations in the critical region.

\subsection{Diffusion coefficient}

In the steady state $\hat{\rho}_{ss}$, temporal fluctuations of the current are captured by the two-time correlation function \cite{wiseman_quantum_2009}
\begin{equation}
\label{2_time_corr}
    F(\tau) =  \lim\limits_{t \to \infty} \{ \mathrm{E}\left[ I(t+\tau) I(t) \right] - J^{2}\}\, ,
\end{equation}
where $J=\mathrm{E}\left[ I(t+\tau)  \right]=\mathrm{E}\left[I(t) \right]$, since the average current is unchanging in the steady state.
Note that $F(\tau) = F(-\tau)$ is an even function of the time difference $\tau$ only, because the steady state is time-homogeneous. By definition, the diffusion coefficient measures the growth of the variance of the integrated current at long times~\cite{esposito_nonequilibrium_2009,Schaller2014}: 
\begin{equation}
    \label{diffusion_coefficient_def}
    \mathcal{D} = \lim_{t\to\infty} \frac{d}{dt}{\rm Var}\left[\int_0^t{\rm d}t' I(t')\right].
\end{equation}
It is straightforward to express this in terms of the autocorrelation function as
\begin{equation}
    \label{diffusion_coeff_autocorrelator}
    \mathcal{D} = 2\int_0^\infty d\tau\, F(\tau).
\end{equation}

In many experimental settings, rather than sampling the temporal correlation function directly, it is simpler to measure the power spectrum
\begin{equation}\label{spectral_density}
    S(\omega) =  \int\limits_{-\infty}^\infty d\tau~e^{-i \omega \tau} F(\tau)\,.
\end{equation}
This describes the amount of power in each frequency component of the measured signal, and is related to the diffusion coefficient $\mathcal{D}$ by
\begin{equation}
\label{eq:Dif}
    \mathcal{D} = S(0).
\end{equation}
Therefore, to assess the critical fluctuations in the measurement currents we must compute the zero-frequency component of the power spectrum for both photodetection and homodyne detection.

\subsection{Photodetection}
\label{sec:photodetection}
% The type of current $I(t)$ stemming from a monitored optical cavity depends on the unraveling used for the master equation. 
% In photodetection, this corresponds to a jump unravelling whereby at each infinitesimal time interval $dt$, the outcome of the measurement is a binary random variable $dN(t) = 0,1$, which has the value $1$ whenever a jump $\hat{\rho} \to  \hat{a} \hat{\rho} \hat{a}^\dagger$ occurs, and zero otherwise. 
% Furthermore, this is a Poisson process where the average jump corresponds to the average rate of jumping $E[dN(t)]=\kappa \tr{\hat{a} \hat{\rho} \hat{a}^\dagger}dt$ and further satisfies $dN(t)^{2} = dN(t)$.
% Thus probability for this jump to occur is $P(dN(t) = 1) = \kappa \tr{\hat{a} \hat{\rho} \hat{a}^\dagger}$, and the the resulting photocurrent is $I(t) = dN(t)/dt$, which will thus be a sum of $\delta$-like peaks. 

We begin by describing how to calculate the power spectrum of the current for photodetection. We first split the Liouvillian $\mathcal{L}(\hat{\rho})$ [Eq.~\eqref{M}] into a free  and a jump evolution respectively~\cite{Schaller2014}:
\begin{equation}
\label{pd_split}
    \mathcal{L} = \mathcal{L}_0 + \mathcal{L}_1, 
    \qquad 
    \mathcal{L}_1(\rho) = \kappa~\hat{a} \hat{\rho} \hat{a}^\dagger. 
\end{equation}
By splitting the Liouvillian into two parts, we can easily define the average current 
\begin{equation}\label{pd_ave}
    \mathrm{E}[I(t)] = \tr{\mathcal{L}_1(\hat{\rho})} = \kappa \langle \hat{a}^\dagger \hat{a} \rangle\,,
\end{equation}
One can also show that the two-point correlation function~\eqref{2_time_corr} is given by~\cite{wiseman_quantum_2009} 
\begin{align}
\label{pd_2time}
    F_{\rm PD}(\tau) &= \tr{ \mathcal{L}_1 \left(e^{\mathcal{L}\tau}\left( \mathcal{L}_1 \hat{\rho}_{ss}\right)\right)} - J_{\rm PD}^2 + J_{\rm PD}~\delta(\tau)\,,
    \\
    &= \kappa^2 \tr{\hat{a}^\dagger \hat{a} e^{\mathcal{L}\tau} \big( \hat{a}\hat{\rho}_{ss} \hat{a}^\dagger)} - J_{\rm PD}^2 + J_{\rm PD}\delta(\tau).
\end{align}
The final Dirac-delta term ensures the two-time correlation function is well defined at $\tau=0$.
Moreover, the first term is related to Glauber's second-order coherence function~\cite{glauber_quantum_1963}, and has a very intuitive interpretation: it is the conditional probability for the system to undergo a jump after a time delay $\tau$ following a previous jump.

Now we seek to find an efficient way of calculating the power spectrum associated with $F_{\rm PD}(\tau)$. To handle this problem, we use a vectorized representation of the Liouvillian~\eqref{M}, whereby we interpret density matrices $\hat{\rho}$ as kets $\ket{\rho}$ in a doubled Hilbert space \cite{landi_non-equilibrium_2021}. Under this transformation, superoperators are interpreted as matrices, and the trace is replaced by a contraction with the vectorized identity $\bra{ 1 }$. That is, for any superoperator $\mathcal{S}$, we may define the trace $\tr{\mathcal{S}(\hat{\rho})} = \bra{ 1} \mathcal{S} \ket{ \rho}$.
We further assume that the steady state is unique and $\mathcal{L}$ is diagonalizable.
% The former assumption excludes for example, degenerate Liouvillians, but the latter incurs no loss of generality: the final results will hold even if $\mathcal{L}$ is not diagonalizable. 
Given that $\mathcal{L}$ is not Hermitian, it will in general have different right and left eigenvector pairs, which we denote as
\begin{equation}
\mathcal{L}|x_j\rangle = \lambda_j \ket{x_j}, 
    \qquad 
    \bra{ y_j}  \mathcal{L} = \lambda_j  \bra{y_j}\,.
\end{equation}
These satisfy the identity $\braket{ y_j }{ x_k }= \delta_{j,k}$. Furthermore, since $\tr{\mathcal{L}\hat{\rho}} = 0$, we know that both the steady state and identity exist in the nullspace of the Liouvillian; i.e., are eigenvectors with a zero eigenvalue, such that $\mathcal{L}\ket{\rho_{\rm ss}} = 0$ and $\bra{ 1 } \mathcal{L} = 0$.
If the system is assumed to be stable---so that it relaxes to a unique steady state---then it necessarily implies that all other eigenvalues $\lambda_j$ have strictly negative real parts, ${\rm Re}(\lambda_j) < 0$.
Bringing all this together, we can write
\begin{align}
\label{L_eigen}
    \mathcal{L} = \sum\limits_j \lambda_j |x_j\rangle\langle y_j |\,,
    \qquad
    e^{\mathcal{L}t} = |\rho\rangle\langle 1 | + \sum\limits_j e^{\lambda_j t} |x_j\rangle\langle y_j |\,.
\end{align}

This representation allows us to recast the average current as $J_{\rm PD} = \langle 1|\mathcal{L}_1 | \rho\rangle$ and the two-time correlation function as
\begin{equation}
\label{F_PD}
    F_{\rm PD}(\tau) =
    \sum\limits_{j} e^{\lambda_j \tau} \langle 1 | \mathcal{L}_1 |x_j \rangle\langle y_j | \mathcal{L}_1 \ket{\rho_{\rm ss}} + J_{\rm PD} \delta(\tau)\,.
\end{equation}
While this formula is appealing, it is difficult to evaluate explicitly since it requires knowledge of the entire spectrum of the Liouvillian, which may be hard to determine, even numerically. 
Instead, we can directly compute the spectral density by taking the Fourier transform in Eq.~\eqref{spectral_density}, yielding
\begin{equation}
\label{S_pd}
    S_{\rm PD}(\omega) = -2  \bra{ 1} \mathcal{L}_1  \left(\frac{\mathcal{L}}{\mathcal{L}^2 + \omega^2}\right) \mathcal{L}_1 \ket{\rho_{\rm ss}} + J_{\rm PD} \,,
\end{equation}
which is equivalent to the power spectrum derived in \cite{brandes_waiting_2008}. Given that $\mathcal{L}$ is singular, but $\mathcal{L}^2 + \omega^2$ is not, the inverse in the above expressions is well defined for all $\omega \neq 0$. Conversely, for $\omega = 0$ some care must be taken. Taking Eq.~\eqref{F_PD} as a starting point, it is clear that 
\begin{equation}\label{pseudo_inverse}
    \lim\limits_{\omega \to 0} \left(\frac{\mathcal{L}}{\mathcal{L}^2 + \omega^2}\right)  = \sum\limits_j \frac{1}{\lambda_j} |x_j\rangle\langle y_j| \equiv \mathcal{L}^{+}, 
\end{equation}
which is the Drazin inverse of $\mathcal{L}$ and satisfies $\mathcal{L}^{+}\mathcal{L} = \mathcal{L}\mathcal{L}^{+} = \mathbb{I}- |\rho\rangle\langle 1|$. 
We thus arrive at 
\begin{equation}
\label{S0_pd}
    \mathcal{D}_{\rm PD} \equiv S_{\rm PD}(0) = -2  \bra{ 1 } \mathcal{L}_1 \mathcal{L}^{+} \mathcal{L}_1 \ket{\rho_{\rm ss}} + J_{\rm PD} ,
\end{equation}
which we can evaluate numerically.

% Using Eq.~(\ref{S0_pd}), we compute the diffusion coefficient for varying parameters detunings $\Delta$, and values of $1/U$. Shown in Fig.~\ref{fig:fluctuations_pd}(a), we can clearly see that at the critical detuning $\Delta_{d}\sim2$, that the fluctuations increase by orders of magnitude for increasing $1/U$ indicating that the divergence is in fact exponential.
% We confirm this by taking the peak value of the diffusion $\mathcal{D}$ for $\Delta > 0$ and plot it as a function of $1/U$ which indeed shows an exponential dependence in Fig.~\ref{fig:fluctuations_pd}(b). 
% We will qualify this divergence more precisely in Section.~\ref{sec:tunnelling rates}
% Finally we can plot the phase space of the diffusion coefficient as $(G, \Delta)$ in Fig.~\ref{fig:fluctuations_pd}(c) which indicates a rapidly increasing along the red-dashed line marking the discontinuous phase transition. 

\begin{figure*}
    \centering
    \includegraphics[width=\textwidth]{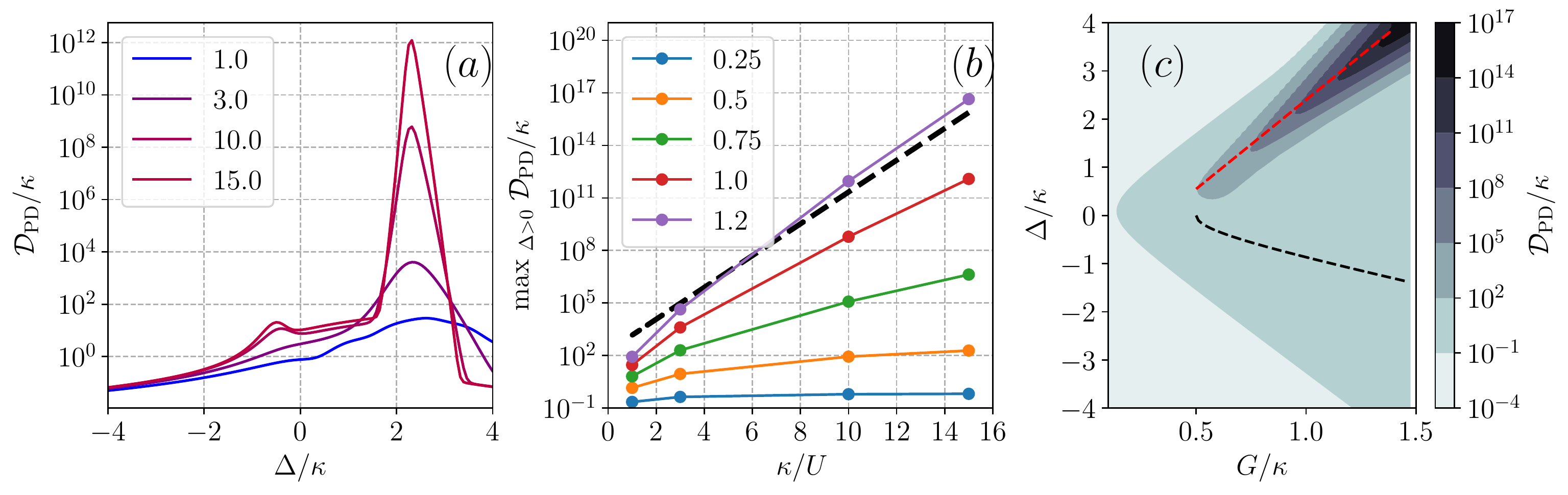}
    \caption{ 
    Diffusion coefficient $\mathcal{D}_{\rm PD}/\kappa$, Eq.~\eqref{S0_pd}, for direct photodetection.
    (a) As a function of $\Delta/\kappa$, for different values of $\kappa/U$ (legend) and fixed $G/\kappa = 1$, showcasing the exponential divergence at $\Delta_{d}>0$ (notice the log-scale).
    (b) Maximum of $\mathcal{D}_{\rm PD}$ over all $\Delta >0$, as a function of $\kappa/U$. 
    Each increasing curve corresponds to the increasing value of $G/\kappa$ (legend).
    Throughout the entire critical region ($G/\kappa > 1/2$) the fluctuations are found to diverge exponentially with $\kappa/U$.
    The black dashed line has slope $\kappa/U$ confirming the prediction that $\mathcal{D}_{\rm PD} \sim e^{\kappa/U}$ in Eq.~(\ref{eq:diff_exp}).
    (c) Diffusion coefficient $\mathcal{D}_{\rm PD}$ in the  $(G,\Delta)$ plane, with $U/\kappa = 1/10$ (compare with Fig.~\ref{fig:phase_diagram}(c)).}
    \label{fig:fluctuations_pd}
\end{figure*}

\subsection{Homodyne detection}
\label{sec:homodyne}

Now we turn to computing the fluctuations in the homodyne current $I_{\rm Hom}(t)$.
Unlike photodetection, we cannot split the Liouvillian Eq.~(\ref{M}) into a jump and free evolution like we did in Eq.~(\ref{pd_split}).
However, we can carefully identify the measurement superoperator in Eq.~(\ref{eq:homodyne})  associated with homodyne detection of the $\hat{q}_\theta$ quadrature:
\begin{equation}
    \mathcal{H}_1(\hat{\rho}) = \sqrt{\kappa} \left(\hat{a} \hat{\rho} e^{-i \theta} + \hat{\rho} \hat{a}^\dagger e^{-i\theta}\right)\,. 
\end{equation}
In this case, the average homodyne current along the given quadrature can be expressed as
\begin{equation}
\label{hom_ave}
    J_{\rm Hom} = \tr{\mathcal{H}_1 \hat{\rho}} = \sqrt{\kappa}\langle \hat{a}e^{-i\theta} + \hat{a}^{\dagger}e^{i\theta}\rangle\,.
\end{equation}
Similarly, the two-point correlation function in the steady state is given by \cite{wiseman_quantum_2009}
\begin{align}
\label{hom_2time}
    F_{\rm Hom}(\tau) &= \tr{ \mathcal{H}_1 \left(e^{\mathcal{L} \tau} \left(\mathcal{H}_1 \hat{\rho}_{ss}\right)\right)} - J_{\rm Hom}^2 + \delta(\tau)\,.
\end{align}
This is structurally similar to the photo-detection result~\eqref{pd_2time}, but involves the operator $\mathcal{H}_1$, which is linear in $a$ and not quadratic like  $\mathcal{L}_1$.  
The first term is now proportional to Glauber's first order coherence function. Note also how the singular term in Eq.~\eqref{hom_2time} is independent of $J_{\rm Hom}$. This is associated to the shot noise of the local oscillator that is used in the homodyne scheme.

Repeating the same vectorization approach used in Sec.~\ref{sec:photodetection}, we arrive at equivalent expressions for the two-time correlation function 
\begin{equation}
\label{eq:F_hom}
    F_{\rm Hom}(\tau) =
    \sum\limits_{j} e^{\lambda_j \tau} \langle 1 | \mathcal{H}_1 |x_j \rangle\langle y_j | \mathcal{H}_1 |\rho\rangle +  \delta(\tau)\,,
\end{equation}
the power spectrum
\begin{equation}
\label{eq:S_hom}
    S_{\rm Hom}(\omega) = -2  \langle 1 | \mathcal{H}_1 \left(\frac{\mathcal{L}}{\mathcal{L}^2 + \omega^2}\right) \mathcal{H}_1 |\rho\rangle  + 1 \,,
\end{equation}
and the diffusion coefficient
\begin{equation}
\label{eq:S0_hom}
     \mathcal{D}_{\rm Hom} \equiv S_{\rm Hom}(0) = -2  \langle 1 | \mathcal{H}_1 \mathcal{L}^{+} \mathcal{H}_1 |\rho\rangle  + 1 .
\end{equation}

\section{Current fluctuations in the critical PPK model}
\label{sec:D_results}

\subsection{Critical divergence of the diffusion coefficient}

We now use the formulas developed in Sec.~\ref{sec: measurements} to analyse the current fluctuations across the critical region of the PPK model. We begin with the measurement current for photodetection, $I_{PD}$. Using Eq.~\eqref{S0_pd}, we compute the diffusion coefficient for various parameters and analyse its scaling with $1/U$. At the critical detuning $\Delta_d$ of the discontinuous transition, Fig.~\ref{fig:fluctuations_pd}(a) illustrates an exponential divergence of the current fluctuations with $1/U$. We confirm this by taking the peak value of the diffusion coefficient $\mathcal{D}_{PD}$ for $\Delta > 0$ and plotting it as a function of $1/U$ in Fig.~\ref{fig:fluctuations_pd}(b), which clearly shows the exponential scaling. 

In contrast, the diffusion coefficient shows a far milder divergence near the continuous transition, visible in Fig.~\ref{fig:fluctuations_pd}(a) as a small hump in $\mathcal{D}_{\rm PD}$ that appears near the critical detuning $\Delta_c\approx -0.9\kappa$ for large $\kappa/U$. In fact, the diffusion coefficient scales algebraically with $\kappa/U$ at the continuous transition. This is in accordance with previous results from classical systems, where current fluctuations have been found to diverge algebraically with volume at a continuous nonequilibrium phase transition~\cite{Nguyen2018}. The diffusion coefficient is plotted across the phase diagram in Fig.~\ref{fig:fluctuations_pd}(c), which highlights that the exponential divergence of $\mathcal{D}_{\rm PD}$ is restricted to the critical line of the discontinuous phase transition.

\begin{figure*}
    \centering
    \includegraphics[width=\textwidth]{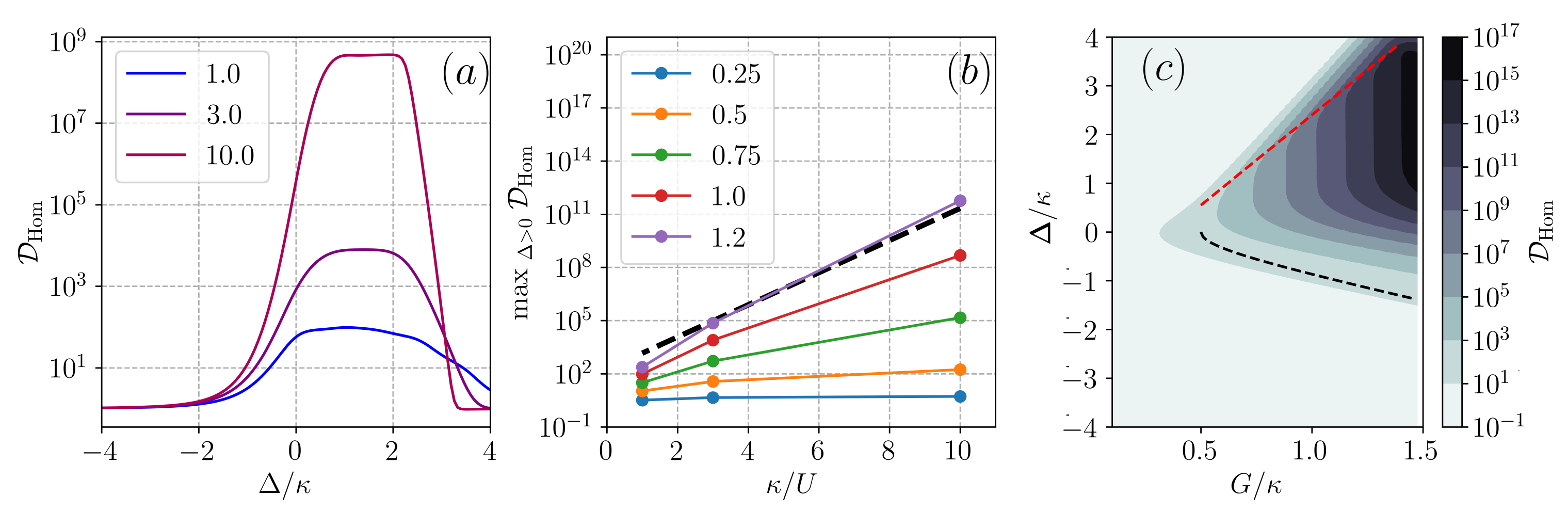}
    \caption{
    This figure uses identical parameters as those used Fig.~\ref{fig:fluctuations_pd}, but for the homodyne diffusion coefficient $\mathcal{D}_{\rm Hom}$ [Eq.~\eqref{eq:S0_hom}]. 
    The exponential divergence is no longer concentrated around the discontinuous transition, but now extends throughout the entire critical region. 
    % (a) The Diffusion coefficient $\mathcal{D}_{\rm Hom}$, Eq.~\eqref{eq:S0_hom}, for the steady-state photon current as a function of $\Delta$, for different $1/U$, with fixed $G = \kappa = 1$.
    % Here we now see we observe an exponentially diverging $\mathcal{D}_{\rm Hom}$ across a range of detunings above the continuous threshold $\Delta > \Delta_{c}$ before a rapid decay when $\Delta > \Delta_{d}$.
    % (b) The largest value of $\mathcal{D}_{\rm Hom}$ for $\Delta >0$, as a function of $1/U$ and different choices of $G$ (with $\kappa = 1$). 
    % Again, the black dashed line is the same as that depicted in Fig.~\ref{fig:fluctuations_pd} and corresponds to the diffusion coefficient when approximated by Eq.~(\ref{eq:diff_exp}) showing an exponential scaling with the semi-classical parameter $n_{0}\propto 1/U$.
    % (c) Here we plot again the $(G,\Delta)$ plane, with $U = 1/10$ and $\kappa = 1$ showing a full region of exponential divergence between both the continuous (blacked dashed) and discontinuous (red dashed) phase transitions. The colour bar corresponds to the values of $\mathcal{D}$.
    }
    \label{fig:fluctuations_homodyne}
\end{figure*}

Turning now to the homodyne current, we find several notable differences as compared with photodetection. As shown in Fig.~\ref{fig:fluctuations_homodyne}(a), the homodyne diffusion coefficient is found to diverge exponentially with $\kappa/U$ over the \textit{entire} critical region $\Delta_c \leq \Delta \leq \Delta_d$. The exponential scaling of $\mathcal{D}_{\rm Hom}$ is confirmed in Fig.~~\ref{fig:fluctuations_homodyne}(b), where the maximum diffusion coefficient is again plotted as a function of $\kappa/U$ for various different values of $G$. Finally, in Fig.~\ref{fig:fluctuations_homodyne}(c) we plot $\mathcal{D}_{\rm Hom}$ in the $(G,\Delta)$ plane, demonstrating that the exponential divergence persists throughout the critical region of the phase diagram, in stark contrast with the photocurrent fluctuations shown in Fig.~\ref{fig:fluctuations_pd}(c). This immediately begs the question as to why the structure of the current fluctuations differs between the measurement processes, and fundamentally what drives the exponential growth in $\mathcal{D}$.

\subsection{Metastability and tunnelling}
\label{sec:tunnelling rates}

% In the Parametrically driven Kerr-model, for large driving and on resonance $\Delta =0$, the state is well approximated by a mixture of positive and negative Cat states---the displacement of which is given by $\alpha$ and the semi-classical solutions Eq.~(\ref{eq:semi-classical})---such that steady state is wekk approximated by a mixture of both positive and negative Cat states
% \begin{align}
% \label{eq:ss}
%     \hat{\rho}_{\rm ss} \approx \frac{1}{2}\left(\ket{\alpha}\bra{\alpha} + \ket{-\alpha}\bra{-\alpha}\right)\,.
% \end{align}

We are now in a position to explain how the exponential divergence in the diffusion coefficient $\mathcal{D}$ originates from tunnelling between metastable states, as illustrated by the quantum trajectories depicted in Figs.~\ref{fig:photodetection} and \ref{fig:homodyne}. Let us first recap why exponentially diverging current fluctuations arise in the classical context, following Refs.~\cite{Nguyen_2020,fiore_current_2021}. Discontinuous phase transitions in nonequilibrium systems are typically characterised by the coexistence of multiple degenerate steady states, which become macroscopically distinguishable in the thermodynamic limit~\cite{Hanggi1984}. In finite-sized systems, {\color{blue}but the transition rates decrease exponentially with volume, so that transitions become rare.}
% transitions between these states can occur albeit with a rate that is exponentially small in the volume. 
These metastable steady states therefore have a lifetime, $\tau_m$, which diverges exponentially in the thermodynamic limit. 
The current fluctuations are sensitive to both fluctuations within each state, as well as the switching between states. 
In fact, as shown in Ref~\cite{fiore_current_2021}, the diffusion coefficient of classical systems, close to the transition point, scale as 
\begin{equation}
    \mathcal{D} \simeq \mathcal{D}_{\rm local} + \frac{1}{4} (J_1-J_0)^2 \tau_m,
\end{equation}
where $\mathcal{D}_{\rm local}$ describes fluctuations within each bistable state, while $J_{1(0)}$ are the average currents in each state. 
Since $\tau_m$ diverges in the thermodynamic limit, we see that as long as $J_1 \neq J_0$, the
diffusion coefficient will be dominated by the last term, and hence will scale as  $\mathcal{D}\sim \tau_m$.

The PPK model is metastable near the discontinuous transition, as witnessed by the telegraphic switching in both the observed currents (Figs.~\ref{fig:photodetection} and~\ref{fig:homodyne}). The dynamics is characterised by two distinct timescales: the first describing fast relaxation within each metastable state, and the second describing rare transitions due to tunnelling between these macroscopically distinct configurations. Each of these configurations, corresponding to the lobes of the Wigner function in Fig.~\ref{fig:wigner1}(e), is effectively a squeezed coherent state. The rate of tunnelling between these states can be estimated from the overlap between the corresponding coherent states~\cite{wielinga_tunneling_1994, wielinga_effect_1995}. In particular, tunnelling between the inner and outer lobes occurs with a rate proportional to $|\braket{0}{\pm\alpha_0}\vert^{2} = e^{-n_{0}}$, where $\alpha_0 = \sqrt{n_0} e^{i\phi_0}$ and $n_0\propto 1/U$ are defined in Eq.~\eqref{eq:semi-classical}. Tunnelling directly between the two outer lobes can also occur but at a rate $|\braket{\alpha_0}{-\alpha_0}\vert^{2} = e^{-2n_{0}}$, which is exponentially smaller. This can be seen in Fig.~\ref{fig:homodyne}(a), where there are many more tunnelling events via the origin than directly between the outer fixed points. Since the lifetime of each metastable state is the inverse of the tunnelling rate, we conclude that
\begin{equation}
\label{eq:diff_exp}
    \mathcal{D} \sim \tau_m \sim e^{n_{0}} = e^{1/U}.
\end{equation}
This explains the exponential divergence in the fluctuations of the observed homodyne and photocurrents, but does not explain the difference between the structure in $\mathcal{D}$ as a function of $\Delta$ when comparing homodyne and photocurrent, as seen in Fig.~\ref{fig:fluctuations_pd}(a) and Fig.~\ref{fig:fluctuations_homodyne}(a), which we  turn to next.

\subsection{Measurement-dependent criticality}

\begin{figure*}
    \centering
    \includegraphics[width=\columnwidth]{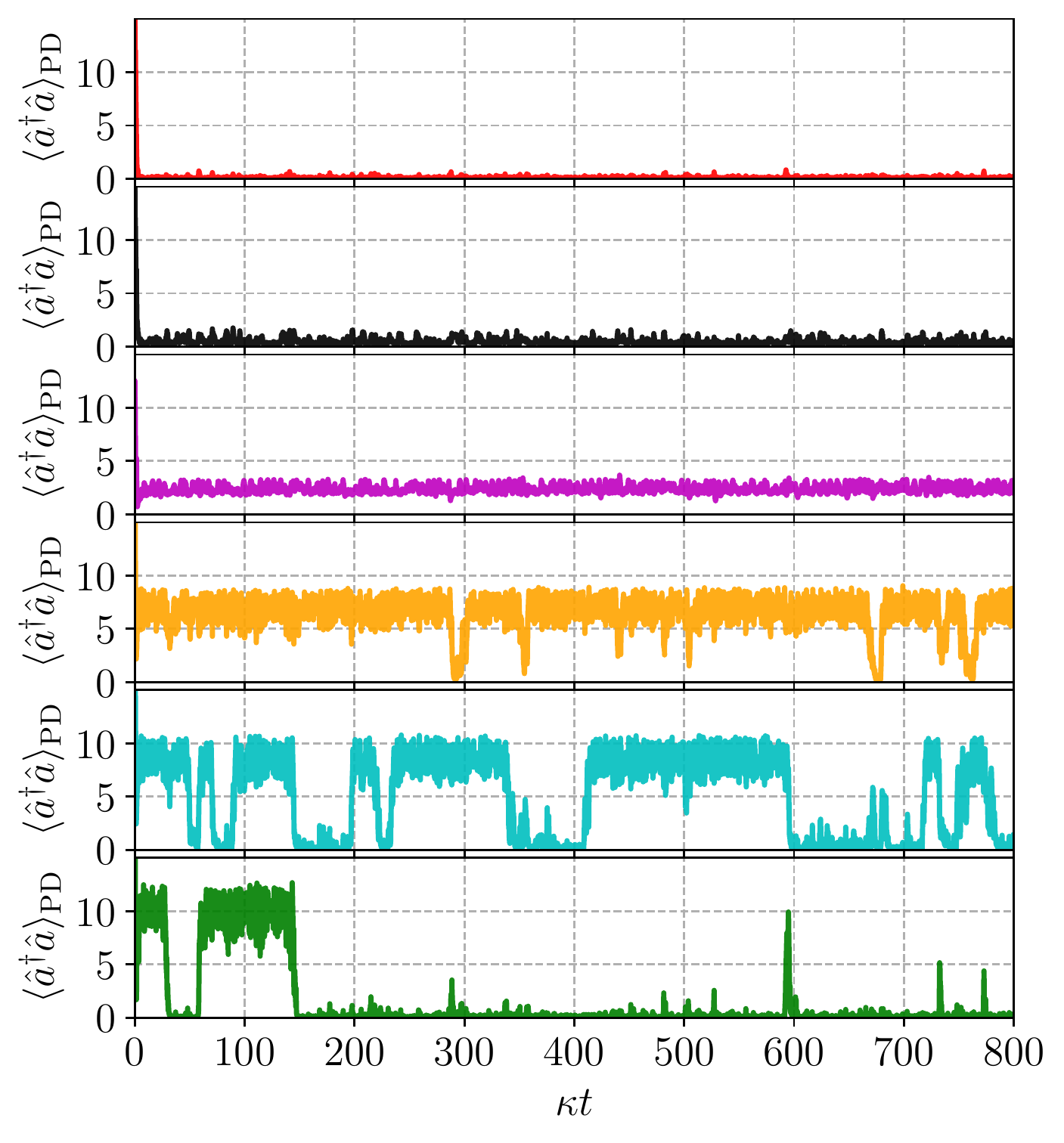}
    \includegraphics[width=\columnwidth]{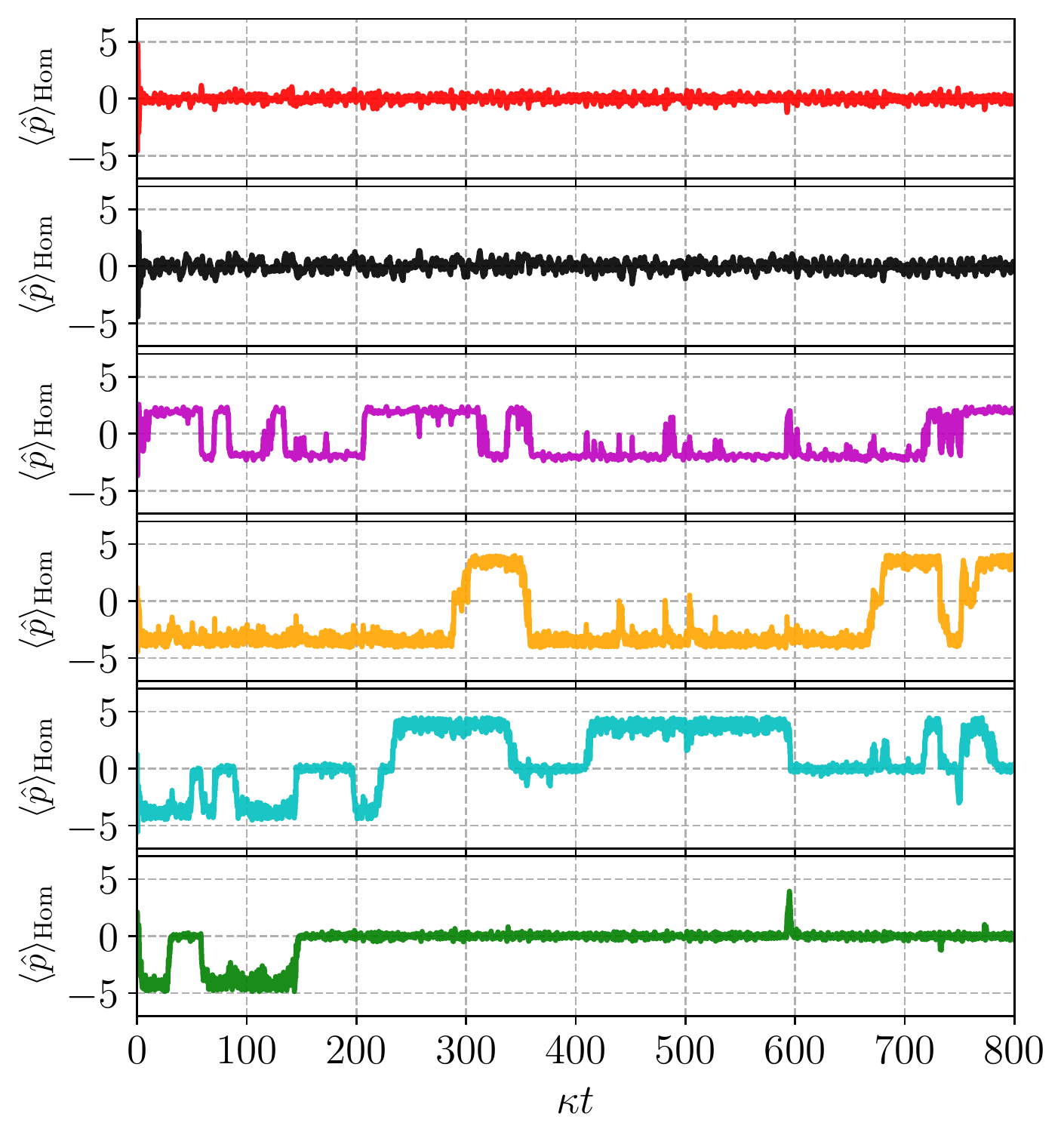}
    \caption{
    Comparison of quantum trajectories for the photodetection mean photon number $\langle \hat{a}^{\dagger}\hat{a}\rangle_{\rm PD}$ (left) and the homodyne phase quadrature $\langle \hat{p} \rangle_{\rm Hom}$ (right).
    Each row corresponds to a configuration in Fig.~\ref{fig:wigner1} from $(a)-(f)$ in descending order with matching colours. 
    When the system crosses the continuous transition, $\Delta>\Delta_{\rm c}$ (third plot from the top), the mean photon number becomes non-zero. But because  photodetection cannot resolve the phase quadratures, tunnelling events cannot be measured. 
    As a consequence, it is not until $\Delta \approx \Delta_{d}$ (fifth from the top) that we see the metastable behaviour appear.
    Conversely, for the case of homodyning (right panel) one can resolve the metastable behaviour already at the continuous  transition (third plot from the top), as they correspond to the  tunneling events between the two lobes in Fig.~\ref{fig:wigner1}(c). 
    This explains why we see an exponential divergence of $\mathcal{D}_{\rm Hom}$  (Fig.~\ref{fig:fluctuations_homodyne}) over a broader region of detunings, as compared to $\mathcal{D}_{\rm PD}$ (Fig.~\ref{fig:fluctuations_pd}). }
    \label{fig:delta_plots}
\end{figure*}

The differing behaviour of the diffusion coefficient for photodetection and homodyne detection can be understood by analysing the quantum trajectories for both measurement schemes. We investigate a series of trajectories, using the same parameters as those used to created the six Wigner functions depicted in Fig.~\ref{fig:wigner1}. The results are shown in Fig.~\ref{fig:delta_plots}, comparing direct photodetection (left panels) to homodyne detection (right panels). 

Under photodetection, as the system initially undergoes the pitchfork bifurcation at $\Delta_c <0$, the only telling signature is a sudden increase in the mean photon number $\langle \hat{a}^{\dagger}\hat{a}\rangle_{\rm PD}$. However, given that photodetection does not contain any information about which fixed point the state is in, the current fluctuates but does not exhibit any signatures of tunnelling between the metastable states. The same cannot be said for homodyne detection along $\langle \hat{p}\rangle_{\rm Hom}$, which immediately collapses the system into one of the metastable states once the system bifurcates. The conditioned moment $\langle \hat{p}\rangle_{\rm Hom}$ then exhibits telegraphic switching due to tunnelling between the two fixed points. As explained in Sec.~\ref{sec:tunnelling rates}, the emergence of this bistable behaviour in the homodyne current $I_{\rm Hom}\propto \langle \hat{p}\rangle_{\rm Hom}$ at the continuous transition explains why we observe exponential growth in $\mathcal{D}_{\rm Hom}$ for detunings far below $\Delta_{d}$ in Fig.~\ref{fig:fluctuations_homodyne}. 

The difference between the two detection schemes is most evident when comparing the purple trajectories (third from the top) in Fig.~\ref{fig:delta_plots}. These corresponding to the Wigner function depicted in Fig.~\ref{fig:wigner1}(c), which has bifurcated along $\hat{p}$ but has not yet reached the discontinuous phase transition at $\Delta_{d}$. We conclude that the distinct behaviour of $\mathcal{D}_{\rm Hom}$ and $\mathcal{D}_{\rm PD}$ is directly related to the type of measurement being made and the information it reveals about the state of the system. This indicates that the choice of measurement basis plays a central role in determining the current fluctuations of quantum critical systems.

\section{Conclusion}
\label{sec:conclusions}

In this article, we have carried out one of the first investigations into the nature of observed current fluctuations in dissipative quantum phase transitions. The system in question is the PPK model, which exhibits both continuous and discontinuous transitions. Both transitions are characterised by the emergence of degenerate metastable states, between which quantum-mechanical tunnelling can occur for finite but large values of $1/U$. When this tunnelling process is resolved by the measurement, the metastability of the NESS reveals itself as exponentially divergent fluctuations quantified by the diffusion coefficient. As we have shown, $\mathcal{D}$ scales with average tunnelling time, which in the PPK model depends exponentially on the mean photon number of the outer fixed points $n_{0}$ and thus explains the observed scaling. Moreover, our results demonstrate that the observed fluctuations intrinsically depend on the choice of measurement scheme, whether it be photodetection or homodyne detection. In particular, since homodyne detection is sensitive to phase it can resolve the tunnelling process near the continuous critical point, unlike photodetection which cannot. 

% Maybe a paragraph here about the Kerr model more generally? Coherent Ising machines etc? Michael would be the guy for this :-) 

While our results pertain to a specific model, our explanation of the diverging current fluctuations is quite general. Therefore, our conclusions should apply to other phase transitions featuring degenerate metastable states, especially in quantum-optical systems that can be continuously measured. For example, much recent theoretical work has focussed on multi-critical behaviour in variants of the Rabi and Dicke models~\cite{Baksic2014,Fan2014,Hwang_2015, Liu2017, Hwang_2018, Peng2019,Soldati2021, Garbe_2020}, in which phase transitions have been experimentally observed~\cite{Baumann2010,Leonard2017b,Cai2021}. The fluctuations of measurement currents in these systems will thus contain a wealth of information about their nonequilibrium phase diagrams.

Looking beyond critical phenomena, current fluctuations have received a lot of attention in the context of stochastic and quantum thermodynamics, with the discovery of thermodynamic uncertainty relations (TURs)~\cite{Barato_2015, Gingrich_2016, Pietzonka_2016, Pietzonka_2017, Pietzonka_2018}. Since TURs connect the fluctuations of currents to the mean entropy production rate, it is unsurprising that nonequilibrium phase transitions---characterised by singular entropy production rates---feature diverging current fluctuations. Nevertheless, the role of TURs in constraining currents in the quantum regime remains the subject of intense research~\cite{Ptaszynski2018,Agarwalla2018,Guarnieri2019}. Our work provides another example where quantum-mechanical current fluctuations can exhibit much richer behaviour than in the classical domain: in this case, because of the different conditional dynamics induced by backaction under distinct measurement schemes. This finding may ultimately prove relevant for ongoing efforts to understand thermodynamic processes under continuous quantum measurement~\cite{Alonso2015,Elouard2017,Naghiloo2018,Naghiloo2020, Belenchia2020,Mitchison2021b,Kewming2022}.

\begin{acknowledgements}
The authors would like to thank P. Potts, M. Brunelli, D. Karevski for fruitful discussions.
GTL acknowledges the financial support of the S\~ao Paulo Funding Agency FAPESP (Grant No.~2019/14072-0.), and the Brazilian funding agency CNPq (Grant No. INCT-IQ 246569/2014-0). 
MJK acknowledges the financial support from the European Research Council Starting Grant ODYSSEY (Grant Agreement No. 758403), and the Marie Sk\l odwoska-Curie Fellowship (Grant No. 101065974).

\end{acknowledgements}

\bibliography{cims}
%\bibliography{references}
% \bibliographystyle{ieeetr}
\bibliographystyle{apsrev4-1}

\end{document}